# Transformative Effects of IoT, Blockchain and Artificial Intelligence on Cloud Computing: Evolution, Vision, Trends and Open Challenges

Sukhpal Singh Gill[1*], Shreshth Tuli[2], Minxian Xu[3], Inderpreet Singh[4,17], Karan Vijay Singh[5,18], Dominic Lindsay[6], Shikhar Tuli[7], Daria Smirnova[6], Manmeet Singh[8,9], Udit Jain[2], Haris Pervaiz[6], Bhanu Sehgal[10], Sukhwinder Singh Kaila[11], Sanjay Misra[12,13], Mohammad Sadegh Aslanpour[14], Harshit Mehta[15,19], Vlado Stankovski[16] and Peter Garraghan[6]

[1]School of Electronic Engineering and Computer Science, Queen Mary University of London, Mile End Rd, Bethnal Green, London E1 4NS, UK
[2]Department of Computer Science and Engineering, Indian Institute of Technology (IIT), Delhi, India
[3]Shenzhen Institutes of Advanced Technology, Chinese Academy of Sciences, China
[4]Department of Computer Science, Simon Fraser University, Canada
[5]Department of Computer Science, University of Waterloo, Canada
[6]School of Computing and Communications, Lancaster University, UK
[7]Department of Electrical Engineering, Indian Institute of Technology (IIT), Delhi, India
[8]Centre for Climate Change Research, Indian Institute of Tropical Meteorology (IITM), Pune, India
[9]Interdisciplinary Programme (IDP) in Climate Studies, Indian Institute of Technology (IIT), Bombay, India
[10]Accenture, Melbourne, Australia
[11]Cvent India Inc, Gurugram, India
[12]Department of Electrical and Information Engineering, Covenant University, Ota, Nigeria
[13]Department of Computer Engineering, Atılım University, Ankara, Turkey
[14]Young Researchers and Elite Club, Jahrom Branch, Islamic Azad University, Jahrom, Iran
[15]Walker Department of Mechanical Engineering, Cockrell School of Engineering, The University of Texas at Austin, Texas, USA
[16]Faculty of Civil and Geodetic Engineering, University of Ljubljana, Ljubljana, Slovenia
[17]1Qbit, Vancouver, Canada
[18]Amazon, Toronto, Canada
[19]Dell Technologies, Austin, TX, USA

s.s.gill@qmul.ac.uk, shreshth.cs116@cse.iitd.ac.in, mx.xu@siat.ac.cn, inderpreet_singh@sfu.ca, kv3singh@uwaterloo.ca, d.lindsay4@lancaster.ac.uk, shikhartuli98@gmail.com, d.smirnova@lancaster.ac.uk, manmeet.cat@tropmet.res.in, udit.cs116@cse.iitd.ac.in, h.b.pervaiz@lancaster.ac.uk, bhanu.sehgal@accenture.com, skaila@cvent.com, sanjay.misra@covenantuniversity.edu.ng, aslanpour.sadegh@jia.ac.ir, harshit.mehta@utexas.edu, vlado.stankovski@fgg.uni-lj.si, p.garraghan@lancaster.ac.uk

[*]Corresponding Author

**Abstract**

Cloud computing plays a critical role in modern society and enables a range of applications from infrastructure to social media. Such system must cope with varying load and evolving usage reflecting societies' interaction and dependency on automated computing systems whilst satisfying Quality of Service (QoS) guarantees. Enabling these systems are a cohort of conceptual technologies, synthesised to meet demand of evolving computing applications. In order to understand current and future challenges of such system, there is a need to identify key technologies enabling future applications. In this study, we aim to explore how three emerging paradigms (Blockchain, IoT and Artificial Intelligence) will influence future cloud computing systems. Further, we identify several technologies driving these paradigms and invite international experts to discuss the current status and future directions of cloud computing. Finally, we proposed a conceptual model for cloud futurology to explore the influence of emerging paradigms and technologies on evolution of cloud computing.

## 1. Introduction

The last two decades have seen active research in the definition and evolution of cloud computing. Driven by innovation in networking and distributed architectures, cloud computing is a manifestation of distributed systems research since the initial conception of the client server model in 1958 [2]. Due to the rapid growth of cloud computing, it has been adopted as an important utility across all aspects of society, from academia, governmental institutions and industry. Characteristics of cloud computing such as dynamic, metered access to a shared pools of computing resources [1] have enabled the realisation of new technologies and paradigms to fulfil the demands of emerging applications including scientific, healthcare, agriculture, smart city, and traffic management [3].

Presently, well-known cloud providers such as Facebook, Google and Amazon utilize large-scale Cloud Data Centers (CDCs) to provision heterogeneous Quality of Service (QoS) requirements. Furthermore, cloud computing platforms are able to provide a unified interface over heterogenous resource found in the Internet of Things (IoT)-based applications which improve the reliability of cloud services [4]. There is a need to sign a Service Level Agreement (SLA) between cloud user and provider to deliver the required service in specified time and budget based on QoS parameters.

Substantial growth in end-user demand and data volume has resulted in the creation of more CDCs at ever growing scale, which in turn increases system energy consumption, $CO_2$ emissions, and waste heat that requires cooling infrastructure for removal. To address the problem of energy consumption, there is a need for new resource scheduling policies to reduce the energy consumption without impacting QoS such as deadline, reliability, availability, cost, security and privacy [5]. To increase the reliability of the cloud computing systems, there is a need to develop new fault tolerant mechanisms, which can maintain the cloud service quality during the occurrence of hardware or software faults. Moreover, security can be improved by using new technology called Blockchain (it is distributed ledgers within Cloud) to protect the communication from attackers, which can further increase reliability of the computing systems [46].

The diversity of large distributed application means there is a requirement for effective big data analytics mechanisms to process the required data in an efficient manner using innovative data processing techniques [61] [66]. Further, new programming models such as serverless computing enable new patterns of resource consumption, autonomically driven by application utilization. Lightweight virtualization provided by container technologies, can improve utilization in clouds, and enable low latency provisioning of application environments [150]. Further, the emergence of fog computing reduces the latency and response time of processing in IoT devices but still research challenges within this domain are not solved effectively. New resource provisioning and scheduling polices are required for fog and cloud computing using Artificial Intelligence (AI) based deep learning techniques to predict the resource requirement in advance for geographically disparate resources [47] [140]. Cloud Computing is emerging as a new tool for solving the complex challenges faced by the Earth Sciences researchers both in the context of compute and analysis [139]. It offers a promising hope for the community which presently relies on dedicated supercomputers coming at a huge cost and can regularly go through slag periods or inefficiencies. The introduction of cloud computing as a supplicant or replacement of dedicated supercomputers is an interesting hypothesis. Due to continuous growing research in the field of cloud computing, there are various new research areas such as quantum computing, software-defined network, software engineering, bitcoin currency, 5G network and beyond are emerged.

## 1.1 Our Contributions

Earlier methodical surveys and system reviews have been identified previous innovations, however innovations in the field of cloud computing require a revisit of paradigms (IoT, AI, and Blockchain) driving cloud computing. There is a requirement for a systematic review to evaluate, upgrade, and integrate the existing research presented in this field with respect to the emerging paradigms and technologies such as IoT, AI and Blockchain. This systematic review presents an updated study to evaluate and discover the research challenges based on the available existing research along with the evolution and history of computing systems as per new frontiers as an amalgamation of these technologies having a high impact on cloud computing and related domains. Finally, we offer critical insights and points out possible future work. We proposed a conceptual model which integrates and enables computation using a plethora of technological advancements and provides an enhanced and holistic setup for next generation computing environments.

The motivation behind this systematic review is to study the history of computing and identify how the emergence of triumvirate *"IoT + AI + Blockchain"* will transform cloud computing to solve complex problems of next generation computing. Further, the international experts of different cloud computing research areas come together and discuss the existing research and proposed future research directions for academicians, practitioners and researchers working in the field of cloud computing. This is the first systematic review which explores the evolution of computing paradigms and technologies and the influence of triumvirate (blockchain, IoT and Artificial Intelligence) to the evolution of cloud computing.

The rest of the article is structured as follows: Section 2 presents the background of cloud computing paradigms and techniques and their evolution. Section 3 presents the drivers (IoT, AI and Blockchain) of cloud computing. Section 4 presents the impact of new paradigms and technologies on cloud computing along with their future research opportunities and open challenges. Section 5 presents the insights of triumvirate to the cloud computing evolution. Section 6 presents a conceptual model for cloud futurology. Finally, Section 7 summarizes the research article.

## 2. Background: History of Decades

Computing systems have evolved from year 1958 to improve the use of hardware resources in an efficient way. During these decades of computing, there have been various types of computing paradigms and technologies have been developed and invented, which contributes extensively to the current research in the field of computing.

### 2.1 Evolution of Computing Paradigms and Technologies: A Journey

Initially, one system can execute one specific task at a time and multiple systems are needed to run parallely to execute multiple tasks concurrently [1]. A secure communication network is required to exchange data among different computing systems. Figure 1 shows the evolution of computing technology along with their objectives and focus of study from year 1958.

- *Client Server*: This is distribution application or centralized system developed in 1960 to divide the workloads or tasks among resource providers (servers) and clients are service requesters [1]. Computer network is used to communicate between servers and clients and server shares resources with clients to execute their workloads in a load balancing manner [2]. Email and world wide web (www) are two important examples of client server model. In this model, clients cannot communicate with each other directly [7].

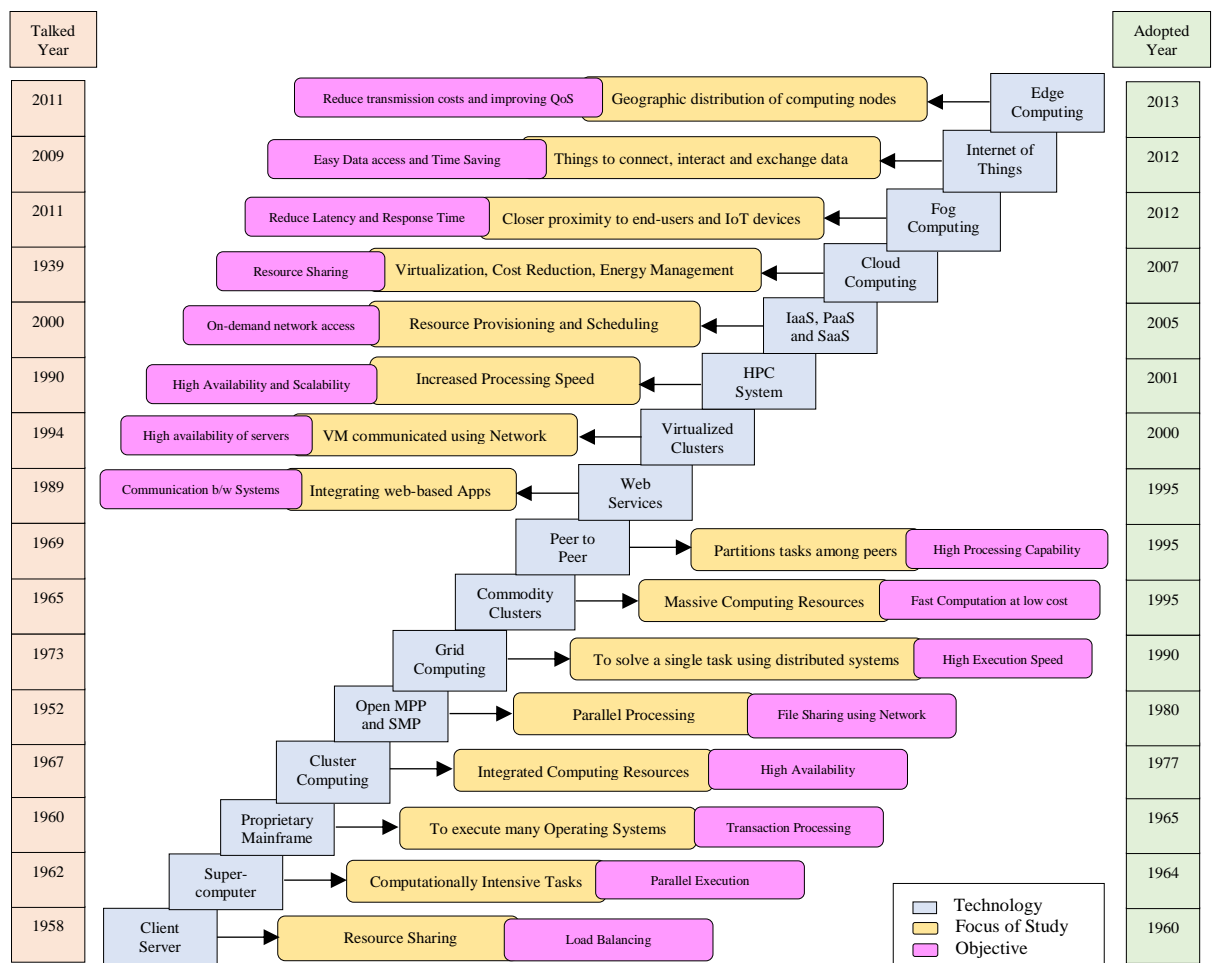

Figure 1: Evolution of Computing Paradigms and Technologies

- *Supercomputer*: It is a system with high performance computing capability to execute computationally intensive tasks in different scientific fields such as molecular modelling, climate research and quantum mechanics [3]. Energy usage and heat management in supercomputers remained a main research challenge thought its evolution since 1960 [7]. The important examples of supercomputers are Multivac, HAL-9000 and The Machine Stops [2].
- *Proprietary Mainframe*: This is large-high speed computer, which can further support various devices and workstations, are used to process large amount of data such as transaction processing, consumer statistics and census [4]. Mainframe computers can provide reliability and security and achieves high throughput through virtualization [7]. In year 2017, IBM launched latest version of mainframe IBM z14 [2]. The performance of mainframe computer is excellent, but these computers are quite expensive.

- *Cluster Computing*: This technology uses fast local area network to communicate available computing nodes and clustering middleware is used to make coordination among different computing nodes [5]. The main objective of cluster computing is to execute a single task using different interconnected computing nodes to improve the performance of computing environment [1].
- *Open Massively Parallel Processing (MPP) and Symmetric Multi-Processing (SMP)*: There are two main types of parallel processing environments: massively parallel processing (MPP) and symmetric multiprocessing (SMP) systems [1] [2] [3]. In SMP environment, other hardware resources (disk space, memory) are shared by multiple processors but using a single operating system. The sharing of resources effects the computing speed of the completion of a particular job [7]. In MPP environment, only file system is shared but there is no sharing of resources during execution of job. The scalability can be improved by adding computers and related disk and memory resources.
- *Grid Computing*: This technology enables to achieve a common objective using distributed computing resources and executes non-interactive workloads which contains huge number of files [4] [5]. The single grid is dedicated to an execution of a specific application [7]. Grid computing provides services such as resource allocation and management service, secure infrastructure and monitoring and discovery service.
- *Commodity Clusters*: It is also called Commodity Cluster Computing, which offers low cost computation of user workloads by using huge numbers of computing resources in a concurrent manner [2] [4]. Different vendors are using open standards to manufacture commodity computers to reduce the variation among products of vendors [7]. Presently, off-the-shelf commodity computers are available to fulfil the business computing requirements quickly.
- *Peer to Peer*: It is a distributed architecture to divide the workload or task among different peers or computing nodes and peers can communicate with each other directly at application layer [4] [7]. In Peer to peer architecture, peers can access different resources such as processing power, disk storage or network bandwidth without the requirement of central coordinator. TCP/IP network is using to exchange data among peers. The main applications of peer to peer architecture are multimedia, file-sharing networks and content delivery.
- *Web Services*: This technology enables the communication among different electronic devices through world wide web using different types of machine-readable file formats such as JavaScript Object Notation (JSON) and Extensible Markup Language (XML) [1] [2]. Basically, web service provides the user interface to end user for the interaction with database server [7].
- *Virtualised Clusters*: It is an implementation of a real computing system to perform similar functions using virtualized environment [1]. Virtualised cluster enables the sharing of resources among different Virtual Machines (VM) to execute workloads or tasks [2]. VM hypervisor provides the software layer-based virtualisation to execute on top of operating system or on the bare metal [7]. VM based computing systems saves resource cost and executes the large number of workloads using same resources.
- *HPC System*: This is the tool which is used to solve large problems (which requires high computing power) of business, engineering and science [5]. High Performance Computing (HPC) system contains different types of computing resources to solve different types of problems and access these resources is controlled by batch system or scheduler [5] [7]. HPC systems are sharing resources and it can access different resources remotely and execute workloads or tasks using scheduling of parallel resources.
- *IaaS, PaaS, SaaS*: There are different types of web services, which can be accessed via Internet such as SaaS (Software as a Service), PaaS (Platform as a Service) and IaaS (Infrastructure as a Service) [6]. SaaS offers software functionality as a service without any maintenance and initial cost with high quality and an example of SaaS is Gmail. PaaS offers the framework, where user can deploy their application with required scalability and an example of PaaS is Microsoft. IaaS offers infrastructure resources such as network, memory, storage and processor to execute workloads or tasks in a cost and time optimized manner and an example of IaaS is Amazon.
- *Cloud Computing*: The cloud services are generally denoted by – XaaS where X = {I, P, S…} and practice of using remote resources to execute user tasks (processing, management and storage of data) through Internet [6]. Cloud computing enables sharing of resources to reduce execution cost and increase availability of service. There are four different types of cloud computing models: public, private, hybrid and community. The Quality of Service (QoS) parameters such as reliability, security and energy efficiency are important to deliver an efficient cloud service.
- *Fog Computing*: This is latest architecture which performs significant amount of storage and computation using end devices or fog nodes and Internet is used to establish communication among these devices [151]. Fog computing comprise of data plane and control plane [6]. Data plane provide services at the edge of network to reduce latency and increase QoS, while control plane is part of router and decides network topology [8]. Further, fog computing supports Internet of Things (IoT) devices such as mobile phones, sensors, health monitoring devices.
- *Internet of Things*: IoT devices are network devices such as actuators, software, home appliances and sensors and Internet connectivity is used to exchange data among these network devices [8]. There are number of applications of IoT in different fields such as agriculture, healthcare, weather forecasting, transportation, smart home and industrial robotics [152] [153].
- *Edge Computing*: It is a distributed computing paradigm, which performs computation on distributed edge devices and it enables the data collection and communication over network [6]. Further, edge computing moves the large volume of data by processing at edge devices instead of cloud server, which improves the QoS, reduce latency and transmission cost [8] [154] [155]. The time sensitive applications can take more advantage from edge computing, but it needs continuous Internet connection to perform dedicated functions within given time.

**3. Triumvirate: IoT + AI + Blockchain**

Cloud computing becomes an intelligent computing with the emergence of innovative technologies and paradigms such as Internet of Things, Blockchain and Artificial Intelligence.

**3.1 Internet of Things (IoT)**

The modern Internet integrates objects known as Things, equipped with sensing, actuating and networking capabilities with dynamic monitoring and control services. Such devices are pervasive in modern life and can be found in homes, public transport, motorways and vehicles [22]. As such, IoT applications are able to operate across heterogeneous domains and enable rich analyses and management of complex interactions [150]. Thus, IoT devices and service are able to address challenges in a wide range of application domains, including e-health, infrastructure, building management systems, manufacturing and transport [23] [24] [25].

The Internet of Things possess several characteristics central to their operation, including (I) Systems are often highly dynamic and network membership must cope with volatility, where a device may appear and reaper across several networks [24], (II) devices are highly heterogenous in terms of both computing performance and functional capabilities, and as such system must cope with limited processing, memory and persistent storage [26], and (III) Systems are managed and controlled by multiple stakeholders, requiring federated mechanisms for secure management of collected IoT data [25].

Historically IoT applications have offloaded processing, and persistent storage to cloud services, however as the number of 'Things' grows, these services fail to support real time demand of IoT devices [24], [27]. This is because such systems operate in physical environments, across large geographic ranges, and as such require low latency response times, and have high density data ingestion requirements/bandwidths [46].

Fog/Edge computing extends cloud system boundaries, by decentralising resource orchestration from datacentres to edge networks [25]. Organised as hierarchal networks of Fog nodes or cloudlets [28] providing deployment of ingestion, processing and management services. Geographic locality allows lower response latencies and increase's ingestion bandwidth by horizontally scaling resources, whilst consuming less energy and enabling resource mobility when compared to cloud services. These characteristics enable IoT applications to scale in terms of a both logical scale and geographic range, whilst providing real-time response latencies, and as such Fog/Edge computing can be considered a future architecture of IoT applications [23].

Smart e-health applications are able to monitor patient data in real-time, by collecting data from implantable and wearable devices forming personal area networks [61]. Smart-Gateways collect and perform local processing of data collected from devices, including noise filtering from medical devices, data compression and fusion, and analyses allowing detection of dangerous trends in a patient's health. Whilst long term trends can be analysed at cloud layers [22], [29]. Furthermore, Fog enabled IoT systems are adaptable and can change their behaviour according to state determined by collected sensors' data. For instance, a smart gateway collecting samples from a pacemaker can increase its sample prior to a heart attack, detected via pre-processing at the fog layer [22], [25], [29].

The Internet of Energy (IoE) paradigm introduces the notion of smart grids and energy management [30]. In which distributed networks energy generators capable of monitoring power consumption and generator, or battery capacity and providing coarse grained statistics about grid health. Whilst 'Smart-Meters' are able to monitor capacity, generation and usage at a finer granularity and report energy demands to utility providers [25], [30], [31]. As such, IoT is enabling technology of future systems, such as electronic vehicles, and micro-grids. Furthermore, such a grid can provide safer, more reliable and robust power delivery, to meet changing consumer demands [32]. The interested readers can further explore using extensive surveys on IoT [8] [22] [24] [25] [26].

**3.2 Blockchain**

Recently Fog, Edge and Cloud computing paradigms have gained significant popularity both in industry and academia. With their increased usage in real-life scenarios; security, privacy and integrity of data in such frameworks has gained high importance [9]. Malicious deletion, theft and corruption of data due to ransomware, trojans and viruses, etc. has been a menace in this domain [10]. Maintaining integrity of data and ensuring that data is not sent by unregistered source are very important for the credibility of the systems [114]. Being used in mission critical applications like hospital care, Smart cities, transportation, surveillance, the tolerance in such systems is very low. As most Edge devices have compute and storage limitations; difficult constraints arise in providing an optimal scheme for data protection and maintaining integrity. To ensure data is protected, Blockchain technology has been adopted in the IoT domain and other real time systems.

Theoretically, Blockchain is a suite of distributed ledgers that can record and track the value of a commodity [11] [129]. Whenever new data is added to the system, it is converted to a Block where a Proof of Work (PoW) is created, which is a hash value difficult to produce without changing the PoW of all blocks preceding it in the ledger. Miners in the Fog system mine the blocks and generate and validate such PoWs. Once a miner completes the proof of work, it publishes the new block in the network and the rest of the network verifies its validity before adding it to the chain. Moreover, this fraudulent manipulation of data in a Blockchain will not be successful unless 50% of its distributed copies are individually reformed by following the same set of operations. Thus, it becomes very hard to alter any data in blockchain within rigid time limit. To support and operate with the blockchain, network peers must provide, the following functionality: routing, storage, wallet services and mining [12].

Despite such problems there have been many efforts to provide robust frameworks that integrate Blockchain with Fog computing [15] [16] [17]. Most such frameworks like FogBus in [15] maintain a dynamically allocated mining strategy where some nodes which are less utilized at a point of time mine and validate the chains and others are used for load distribution, compute and data collection. If any worker reports error in terms of Blockchain tampering or signature forgery, then the Blockchain in majority of the network is copied to that node. Other additional features that they provide are encryption with dynamic exchange of public key pairs for identity authentication.

Although, the key concept of Blockchain is simple, it faces from several challenges while integrating with Fog computing frameworks. Storage capacity and scalability are highly debated due to high cost and maintenance overheads [13]. Even though, only full nodes (nodes that can fully validate the transactions or blocks) store the full chain, the storage requirements are significant. Another weakness of blockchain is the anonymity and privacy of data. Privacy as such is not embedded in the design of blockchain and hence third-party software are required to achieve this. This may lead to unoptimized implementations which are more expensive in terms of compute and storage requirements.

Many open challenges and directions are still existing where blockchains can be improved in IoT frameworks. Resource limitation is the main limiting factor for high quality data protection and reliability. Due to resource constraints, highly sophisticated encryption or key generation cannot be integrated with such chains of data. Only limited cryptographic algorithms can be deployed. More efficient algorithms can be developed keeping in mind of the resource constraints. Another important direction is the modification of such chains in high fault rate settings where the edge nodes can be compromised at any instant of times. Revalidating blocks and copying chains from the majority network leads to large overhead on network and I/O bandwidth requirements. Most frameworks also work in a Master-Slave fashion and hence have single point of failure. This is natural in heterogenous environments. Significant research is needed to ensure redundancy while keeping in view the costs and reliability trade-offs. Also, the blockchain vulnerabilities still affect Fog frameworks [18]. Effective consensus mechanisms need to be developed that can validate blocks with limited sharing and copying of blocks. The interested readers can further explore using extensive survey on Blockchain [179].

### 3.3 Artificial Intelligence

Artificial Intelligence (AI) aims to make IoT and Fog nodes aware of the workload environment and continuously adapt to provide better QoS characteristics, reduce power consumption or overall cost of the infrastructure. AI encompasses various search algorithms, machine learning, reinforcement learning and planning [146]. In the modern world of data intensive tasks with growing fog and cloud deployments, more and more intelligence are required at different levels to provide optimum task scheduling decisions, VM migrations, etc. to optimize mentioned previously under various constraints. These constraints can range from computation capabilities, bandwidth limits to SLA or deadline requirements of tasks.

There have been several works that aim at leveraging AI techniques to improve the performance of fog and cloud systems [144][145][147][148][149]. Different works focus on optimum scheduling policies for cloud, virtualization algorithms, distribution systems among others. They use search methods like genetic algorithms, supervised machine learning and even deep reinforcement learning to optimize their objective functions [154] [155]. AI provides a lucrative avenue to optimize large systems with huge amounts of data with engineering simplicity and efficiency by allowing automated decision making instead of human encoded heuristics which provide more efficient decisions very quickly.

Cloud computing is growing quickly, and CDCs become an important part of eminent industries such as Facebook, Microsoft, Google, Amazon [58]. However, it is difficult to monitor the performance of large-scale cloud data centres manually. Yotascale is a next generation computing and automatic performance monitor solution to reduce the accountability on humans. Yotascale uses historical data to make forward predictions or decisions about cloud costs using Artificial Intelligence and helps to save more cost. Further, real-time analysis can be done using Yotascale to detect anomalous trends using deep learning techniques (supervised/unsupervised techniques or prediction models), finds the root cause and gives future predictions about cloud usage and its cost. The interested readers can further explore using extensive surveys on AI [180] [181].

### 4. Impact of New Paradigms and Technologies on Cloud Computing: Open Challenges and Trends

Cloud computing is evolving very rapidly, and various number of researchers and academicians are working actively to solve the research challenges existing within the cloud computing domain. We have identified various research areas related to cloud computing, which utilizes its available technologies and paradigms in an efficient manner to solve the current problems. Figure 2 shows the emerging research areas for future practitioners, industries, academicians and researchers.

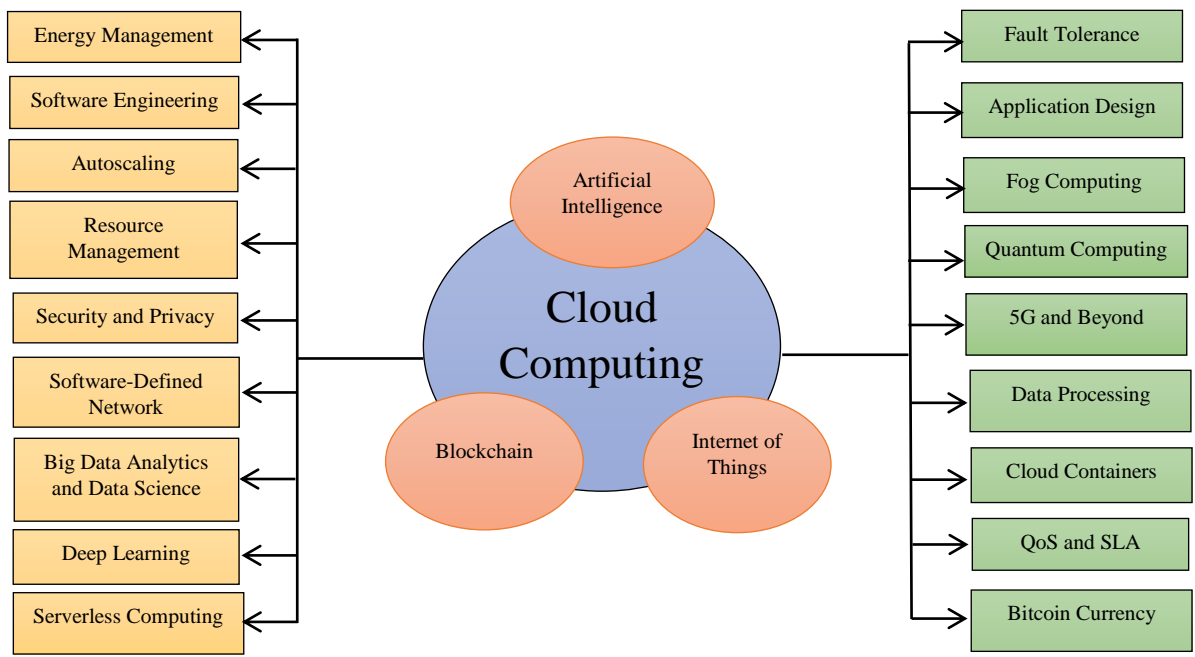

Figure 2: Emerging Research Areas

We have given the foundations of emerging paradigms and technologies for researchers, academicians and practitioners in this section and their corresponding future research directions and open challenges are presented at the end of every subsection.

**4.1 QoS and SLA**

Quality of Service (QoS) is an important challenge of cloud computing systems, which can predict the performance of the system at runtime [6]. The QoS parameters such as execution time, cost, scalability, elasticity, latency, reliability etc can measures the performance of the computing system [46]. QoS parameters are defined using Service Level Agreement (SLA), which is an official document signed between cloud user and provider based on different QoS parameters [157]. In this era, there are wide range of IoT applications which have different QoS parameters based on their domain, purpose and requirement [52] with more stringent security requirement which can utilize blockchain and related technologies. Further, SLA can be measured using a metric called SLA violation rate, which can estimate the deviation of actual SLA from the required (estimated/predicted) and decides about the compensation in case of SLA violation [53]. QoS is progressively important when comprising cloud services because damaging QoS in one of the services can hazardously affect the QoS of whole computing system. To offer an efficient cloud service, there is a need to provision the required amount of cloud resources, which can fulfil the QoS of an application such as budget, response time and deadline [55] [56]. Therefore, cloud providers must ensure to provision enough resources to avoid or reduce the SLA violation rate in order to execute the user workloads within their specified deadline and budget. The future of next-generation computing systems depends on the QoS-aware resource management mechanisms, which can identify and satisfy the QoS requirements of the computing system [6] [58].

There are various research challenges [59] [61] [63] [66] [67] are existing which hinders the achievement of QoS in an efficient manner. Firstly, there is an unavailability of cloud resources to execute an application at runtime, which increases the execution time and reduce the performance of the system. Secondly, there is a need of an efficient SLA-aware resource management mechanisms reduces the SLA violation rate and maintain the performance of the computing system. Thirdly, there are different SLA standards for the different cloud providers and there is a need of centralized SLA standard to achieve the common goal of multi-cloud environment. Finally, there is a need to find out the trade-off among different QoS requirements due to wide range of IoT applications are running on cloud computing systems using AI based supervised/unsupervised learning techniques or prediction models.

*Open Challenges and Trends:* The interested readers can further explore using extensive survey on QoS and SLA [166]. The open challenges and future research directions for *QoS and SLA* are summarized as follows:

1. There is a need to find out the trade-off among different QoS requirements due to wide range of **IoT** applications are running on cloud computing systems.
2. Applications should be able to provide optimum QoS and SLA characteristics with minimal overhead for maintaining data integrity with **blockchain.**
3. **AI** based approaches have entirely changed the landscape of IoT applications, also the portable devices for transmitting multimedia content in IoT application has become very necessary for the end-users.

**4.2 Fog Computing**

Fog computing is the form of distributed computing paradigm which acts as a middle layer between IoT devices and the cloud data centers [150]. Fog is supplement to the cloud, not a replacement. Fog can never replace cloud completely as we still need cloud to handle big or complex data problem. Fog just provides more or less services which cloud does not provide in the same time limit or latency requirements [151]. Fog reduces latency so the applications which need low response time (as real -time applications i.e., traffic system, emergency system, healthcare system etc.) go for fog services instead of cloud services. However, it fails when the data is very heavy to process by its local nodes [46]. If the application is not time sensitive, then use of computing services of cloud datacentre is more beneficial. The term cloudlet is used to denote small data centres, which have almost the same types of features as the large data center, but in small capacity, like they can process small data not very large data [61]. They are close to the site of data production. Cloudlet can be single computer or group of computers with Internet connectivity. Cloudlets reduce the incoming bandwidth required by the centralized data center. It was becoming very difficult to timely handle large amount of data in centralized cloud environment. Huge data production by IoT devices is also the reason to accept fog computing paradigm as these data cannot be handled by the current cloud system even with high computation ability [128]. Apart from low latency, various services like mobility support, security, good performance and low bandwidth requirement are also provided by the fog. Fog generally provides a resource rich middle layer between the end devices and the cloud system to meet the above objectives. It acts as a bridge between end users and the cloud system. Fog nodes are also attached with the cloud with the help of Internet to use its computing power and storage services. Fog nodes generally analyses the data produced by these end devices and sensors [152]. Then the important output data is sent to cloud so that further processing and storage can be taken place. Therefore, the concept of edge computing came into existence. It provides the computing service at the edge of network [129]. Many applications need very less response time, it is not possible to handle them at very far located data center and reply them back on time. Thus, small data center, cloudlet with little processing power are established at the edge or near data generation site. They can handle these applications in time.

Possible areas where this type of system is needed such as healthcare system [61] [153], traffic system, emergency system etc. But the problem with technology is that it needs more complex resource management strategy. Cloudlet also faces some issues which are as following: - 1) VM handling services must be easily transferable from one cloudlet to another cloudlet as these devices are mobile, they also need to switch among cloudlets for easy operation. 2) Since, there can be any number of cloudlets near end user devices, so there must be a policy to first search, choose and then join with the best cloudlet from many cloudlets before the provisioning. 3) Cloudlets are needed to be more efficient in provisioning as they are connected with the mobile devices. Thus, the behaviour of these devices is quite dynamic as they are mobile. In IoT applications, every simple entity can also become part of it using a device like Radio Frequency IDentification (RFID) tag etc. That can act like data connecting device which will connect these simple devices via Internet. Number of devices connected to IoT is increasing exponentially. This cannot be handled by IoT alone, therefore, amalgamation with fog and cloud is done to perform the action efficiently which reduces latency and response time. Moreover, security threats on such devices have risen in recent years leading to increased requirements of security and data integrity technologies like blockchain.

As this amalgamation becomes more popular different challenges arise. With the rise in compute requirements, more and more tasks need to be executed with better QoS. The scheduling of these applications is complex due to several factors. Firstly, due to the heterogeneity of the computation resources and the hierarchy of fog nodes, there is significant differences in the compute capabilities, speed, response time and energy consumption between edge and cloud resources. Moreover, the mobility of the IoT devices changes the response time and bandwidth capacities dynamically. Furthermore, the tasks are stochastic in nature of their arrival, expected completion times, QoS requirements which makes this problem too complex. Most existing job scheduling algorithms are based on heuristics [141] [142] [143]. Other works use adaptive techniques to optimize job placement and migration decisions [144] [145]. Still, the current works focus on the aspect of scheduling with a limited perspective.

It is well known that heuristics work for generic cases and fail to respond to the dynamic changes to the environments. The adaptive schedulers still lack the ability to optimize for a diverse set of user or application requirements and no model or scheduling architecture exists which aims to optimize multiple objectives simultaneously. Some works focus on optimizing energy, other focus on response time or SLA violations. However, for the diverse needs of users, there is a requirement of schedulers which can optimize multiple metrics at once, prioritizing those that the application needs. There can be a convex combination of multiple such metrics and higher weight can be given to those that are of primary importance. For example, mission critical applications like surveillance, healthcare can have an optimization object with higher weight to response time. Scientific applications can give higher weight to quality of results. Energy sensitive applications like smart cities can have an all-round objective with more weight to energy. Similarly, for other applications such functions can be modified. In this regard, AI based techniques can be leveraged to provide more efficient and robust algorithms for enhanced and user requirement directed task placement.

*Open Challenges and Trends:* The interested readers can further explore using extensive survey on Fog Computing [167]. The open challenges and future research directions for *Fog Computing* are summarized as follows:

1. Generic interfaces are required for fog gateways to be able interact with the plethora of **IoT** devices.
2. **Blockchain** APIs must be runnable on resource constraint fog/edge nodes.
3. State of the art **AI** techniques can be used for proper task scheduling on heterogenous fog environments.

## 4.3 Energy Management

The amount of data collected and processed has been increased manifold in the past few decades. This trend has been forcing the computation and thus the power consumption capabilities of Cloud platforms to extremes. There has been an increase in the electricity consumption of cloud datacentres by about 20% to 25% every year [72]. Due to this, a turn has been observed towards distributed computing, resulting in increasing popularity of Fog and Edge Computing platforms. The shift from centralized Cloud-based computing to edge devices and networks immensely helps in reducing latency [73], improving Spectral Efficiency (SE) and increasing cost effectivity.

However, this comes with its challenges. In many mission critical and remote sensing applications, intermittent power supply, if not the power delivery itself poses serious challenges. With the massively growing number of IoT devices [74] and the increasing data collected, the data-handling capacity, computation capability and bandwidth requirements of networks are being pushed to its limits. On the other hand, smaller IoT devices with low computation power, storage and battery are being developed [128]. Thus, increasing the efficiency of Fog/Edge nodes has become crucial. At the same time, maintaining the sustainability of the Cloud datacentres and reducing carbon footprint [75] [76] has also gained importance. All of this has to be done without compromising on the Quality of Service (QoS) [6].

Despite the challenges, several developments can be seen in this field. Energy management has been tackled in three major levels, namely software, hardware and intermediate. Software-level efficiency optimization techniques and algorithms (like Mobile Edge-Computing offloading in [77]), backed by simulation models [78] have been developed. On the hardware side, application specific devices have been developed to deliver high performance and reduce the power footprint [79]. Wireless Sensor Networks (WSNs) have discussed energy conservation in detail [80] [81]. At the intermediate level, resource management and active Fog/Edge-node sleep duration scheduling techniques and other energy conservation architectures [10] have been employed. For sustainable Cloud Computing, a comprehensive taxonomy has been proposed [58].

Many open challenges and directions for improvement remain where efficiency and sustainability of Fog/Edge/Cloud platforms can be improved [47]. More sophisticated algorithms need to be developed to encode information into a smaller number of bits to reduce the bandwidth budget and thus the transceiver's power requirement (which is much more significant than the CPU itself). Common encoders in almost every mobile device can be exploited to employ encoding techniques without the requirement for extra/dedicated hardware. However, an increasing amount of data sharing, and loss has made it difficult to reduce the theoretical bandwidth. Another direction is to develop thermal-aware resource scheduling for reduced heating, thus improving efficiency. With emerging 3D SoCs and memories, CPU and data usage planning need to be modelled at the transistor level, using 3D thermal simulation architectures developed. Finally, the goal is to reduce power consumption to bare minimum, so that energy harnessing/scavenging techniques can be used to power both the CPU and the transceiver, making the node a completely independent entity. Therefore, reduction in the granularity of the Fog/Edge network can be achieved resulting in widely dispersed, redundant and more fault tolerant frameworks [139]. Other domains like energy constrained blockchain models can be explored with other adaptive AI based learning models for more efficient energy scheduling.

*Open Challenges and Trends:* The interested readers can further explore using extensive survey on Energy Management [58]. The open challenges and future research directions for *Energy Management* are summarized as follows:

1. Enhanced algorithms for efficient data encoding for reduced bandwidth consumption and energy efficient communication in data intensive **IoT** devices.
2. **Blockchain** design should allow energy constraint execution.
3. Using novel **AI** motivated techniques for more efficient thermal aware scheduling of tasks and resources.

## 4.4 Resource Management

Resource management in distributed systems is a challenging task due to the scale of modern data centres. The diverse nature of network devices, components and communication technologies in large-scale distributed systems makes the complexity of resource management techniques increase [52]. Therefore, there is a demand for new resource allocation approaches that would contribute to stability and efficiency of such systems. Resource management is a core concept within distributed systems (including Cloud, IoT, Fog computing), however there must be assurances that such systems exhibit high performance, latency-sensitivity, reliability, and energy-efficiency [46] [47]. These systems do not simply comprise the software layer, but must also factor in other systems including networking, server architecture, and even cooling. The security of cloud system can be increased by utilizing the blockchain technology during the sharing of resources or VM migration.

There is a need to explore new techniques for resource management for computing systems by considering a holistic view of the system by utilizing AI techniques [47]. Further, experiment driven approaches can be explored to investigate techniques to optimize resource management approaches. There is a need to incorporate data abstraction in resource management and its one example of a cluster management system is Borg [58]. This system hides details about resource management so that users can focus on development of applications. Borg logically partitions the whole cluster into cells, each one containing a Borgmaster (controller) and a Borglet that starts and stops tasks in a cell [5]. The master handles client Remote Procedure Calls (RPCs) that can request to create a job or to read data, and it also communicates with the Borglets. This is a very scalable centralized architecture. The key design feature is that even if a master or a Borglet goes down, the already launched tasks will keep running.

In order to enable fair sharing of commodity clusters, a platform called Mesos can be used [6]. It manages sharing of commodity clusters between different frameworks that run on these clusters. The main principle is using resource offers. Mesos decides how many resources to allocate to each framework based on framework's constraints, while those in turn decide which offers to accept. Therefore, the burden of making scheduling decisions falls on frameworks. Besides, Mesos allows the development of specialized frameworks (such as Spark) that could significantly improve performance. A framework called YARN is used to perform resource management and scheduling [7]. It allows applications to request resources on different levels of topology – machines, nodes, racks etc. YARN resource manager is the main component responsible for making allocation decisions [63]. Just like Mesos, it lets commodity clusters to be shared among many frameworks. YARN also has built-in fault tolerance that hides the complexity of fault detection and recovery from its users [130].

*Open Challenges and Trends:* The interested readers can further explore using extensive survey on Resource Management [6] [52] [56]. The open challenges and future research directions for *Resource Management* are summarized as follows:

1. QoS-aware autonomic resource management is required to run IoT based applications without violation of SLA at runtime.
2. Proper **blockchain** mining and hash generation allocation must be done for load balanced execution.
3. New resource provisioning and scheduling polices are required for fog and cloud computing using **AI** based deep learning techniques to predict the resource requirement in advance for geographically disparate resources.

**4.5 Fault Tolerance**

Cloud provider should provide the continuous service to users while maintaining the reliability of the cloud services even in the presence of faults [63]. There is a one mechanism called fault tolerance, which is used to provide the service in an efficient manner while satisfying the QoS requirements of the computing system. The faults occur during the working of computing system can be software, hardware or network. Further, the fault tolerance ensures the robustness and availability of cloud services [47]. The other issues related to reliability in cloud computing are timeout failures, overflow failures and resource missing failures. The other failures can be generated by catastrophic failures, which often leads to cascading systems failures. There are various proactive and reactive fault tolerance techniques are proposed to deal with such kind of failures [63]. Checkpointing is the most popular fault tolerance technique, which is used for long running process by saving the states after every change. Further, checkpoint is used when there would be any failure to start from the same point. Another renown technique is replication-based fault tolerance, which replicates the nodes or tasks to finish the job within their required deadline [58]. Task migration-based fault tolerance technique can migrate job to another machine if current machine is busy or suffering from some failures. To maintain the reliability of the computing systems, the existing fault tolerance techniques need failure-aware provisioning models, autonomic reliability-aware resource management technique and service reliability mechanisms, trustworthy data integrity (blockchain) [52].

Reliability in cloud computing makes an impact on QoS while delivering the cloud service in an efficient manner. One of the most important challenge in cloud computing is how to provide an efficient and reliable cloud service while reducing the energy consumption as well as carbon footprints [58]. There is a need of Reliability-aware cloud as a service to offer resilience with required QoS and system performance [6]. Further, an efficient resource management needs to consider different failure and workload models to execute different types of applications such healthcare, smart city and agriculture [46]. Failure prediction in cloud computing systems is also a challenging task and which can also affect the reliability of the system [63]. There is a need to consider various machine or deep learning techniques [128] [129] to predict the failures and achieve the required reliability of the cloud service to maintain the QoS. For IoT applications, the replication-based fault tolerant techniques are efficient, which can improve the latency and response time of task. Further, to handle the big data applications, there is a need of reliable cloud storage system to provide an efficient retrieval system for processing of big data.

*Open Challenges and Trends:* The interested readers can further explore using extensive survey on Fault Tolerance [63]. The open challenges and future research directions for *Fault Tolerance* are summarized as follows:

1. For **IoT** applications, the replication-based fault tolerant techniques are efficient, which can improve the latency and response time of task.
2. An analytical modelling framework for Practical Byzantine Fault Tolerance (PBFT)-a consensus method for **blockchain** in IoT networks is required to define the viable area for the wireless PBFT networks which guarantees the minimum number of replica nodes required for achieving the protocol's safety and liveliness.
3. There is a need to consider various **machine or deep learning** techniques to predict the failures and achieve the required reliability of the cloud service to maintain the QoS.

## 4.6 Security and Privacy

Recently, in research and industry, there has been a massive shift from personal computing to IoT, Edge and Cloud computing to provide smarter and more efficient services to end users. For this big shift in paradigm, many issues and challenges have arisen in the privacy and security pertaining to the data on these devices. Due to various characteristics of Edge computing like low latency, geographic distribution, mobility of end device, and high processing, heterogeneity, etc [58] [128]. The security and privacy properties need to be more robust and versatile. Moreover, the diversity of applications and heterogeneity of devices makes it difficult to develop seamlessly connected software platforms. To study these security and related concerns in cloud and fog computing paradigm, the following factors become prominent: (1) Trust and privacy of end users (2) Internode source authentication and validation (3) Impenetrable communications among, sensors, compute and broker nodes (4) Identification and protection of the systems against malicious attacks (5) Robust data management and tamper proof databases (blockchain) [93].

Existing work in this area focus to solve challenges like detection and recovery from malicious or malfunctioning nodes, identification of and safeguard against attacks, prevention of malicious threats, safeguarding user-information against theft, dynamic mutual authentication [95] [96]. Recent work has made possible to identity and location privacy for Unmanned Aerial Vehicle (UAV) assisted compute nodes, keeping in mind their integration in the distributed frameworks [97]. Also, in Fog forensics, other works have provided digital evidence by reconstructing past computing events and identified how the key characteristics are different from cloud forensics [94]. Mobility management, interference mitigation, and resource optimization in Fog Radio Access Networks (F-RAN's) [98] are some of the main topics which have had many contributions in recent past. New models have emerged for diverse applications addressing privacy issues. Some such directions include face identification and resolution, vehicular crowd-sensing, geo-location sensing and data analysis, storage architectures & data centres with renewable nodes, fog based public cloud computing [82], [83] [92], [99], [100]. Such works have addressed concerns regarding many vulnerabilities including protection against data theft, man in the middle attacks, user anonymity, location privacy, forward secrecy, secure user level key-management, among others [87].

Many of the privacy and security models developed for fog/cloud computing face some limitations in terms of their scalability to the next generation edge computing shift [88]. Due to the inherently decentralized nature of fog computing, many unforeseen security threats arise in the fog layer and IoT devices which are not a concern in cloud computing [84] [85] [86]. Threats to edge focused networks include Advanced Persistent Threats (APT attacks), threats caused by bi-directional communication, malware, Distributed Denial of Service (DDoS) attacks, micro servers lacking hardware protection mechanisms in edge data centres, restricting the authentication protocols that can be deployed [89] [90] [91]. These works also highlight future directions in Mobile Edge computing framework including high speed pertaining to real-time encryption using nodal collaboration of edge networks.

In prior works, security issues are exploited from a narrow perspective and computing capabilities of both edge and remote resources have not been fully leveraged [85]. Once cloud computing-like capabilities are brought to the edge of the network, novel situations arise. Some such situations include collaboration between heterogeneous edge data centres, migrating services at a local and global scale, concurrence to end users, quality of services, real-time applications, load balancing, server overflow problems, detection of stolen devices, robust and reliable inter node communication. These are the avenues for future research. To solve these problems, other domain ideas can be explored like clustering model-based security analysis (AI based prediction models) which is useful in DDoS attack mitigation in server systems and in intrusion detection systems, evolutionary game theoretic approaches to the privacy models inspired by the adversarial attack models, communication protocols in Sensor cloud systems. The security mechanisms also need to consider the existence of mobile devices using these data-centres.

*Open Challenges and Trends:* The interested readers can further explore using extensive survey on Security and Privacy [168]. The open challenges and future research directions for *Security and Privacy* are summarized as follows:

1. Due to the inherently decentralized nature of fog computing, many unforeseen security threats arise in the fog layer and **IoT** devices which are not a concern in cloud computing.
2. With integrity, **blockchain** structures should also allow other security measures for data like encryption, signature management, etc.
3. **AI** based security-aware techniques can be explored like clustering model-based security analysis which is useful in DDoS attack mitigation in server systems and in intrusion detection systems.

### 4.7 Software-Defined Network

There is a need to enable the concept of networking virtualization in cloud is called Software-Defined Network (SDN) and utilize SDN for cloud computing by extending the idea of virtualization of all the cloud resources such as network, storage and compute [131]. Further, it improves the abstraction of physical resources and automation and optimization of configuration process. SDN paradigm provides a platform to enable flexibility or agility in network, which can be create a cost-effective communication among modern cloud datacenters. Further, SDN based cloud computing reduces the power consumption while optimizing the network virtualization.

Network functions virtualization (NFV) is another emerging networking paradigm which forwards network functions such as Domain Name Service (DNS), load balancing and intrusion detection while executing software-based applications [132]. Moreover, NFV improves the elasticity of network function and increases the flexibility and agility of the service, which further reduces the cost [133]. Further, an efficient VM migration policy can be used for VM consolidation in virtualized network to reduce energy consumption. There are different research challenges are still open for academicians and researchers. Firstly, there is a need provide the security mechanism for SDN-based cloud computing to secure the transfer of data among different cloud datacenters [134]. Adel et al. [116] developed a low-cost Raspberry-Pi-based micro datacenter for software defined cloud computing, which saves cost, but reliability of service is still questionable. Secondly, the trade-off between cost and energy consumption is still existing due to replication of SDN enabled cloud infrastructures. In future, there is a need to deploy SDN-based cloud computing environment, which can reduce energy consumption and increase reliability while providing the network virtualization service in a cost-effective manner. Further, we can extend existing data integration in such SDN environments to support blockchain technologies with enhanced data distribution and results collection techniques motivated from AI based models.

*Open Challenges and Trends:* The interested readers can further explore using extensive survey on Software-Defined Network [131]. The open challenges and future research directions for *Software-Defined Network* are summarized as follows:

1. There is a need provide the security mechanism for SDN-based cloud computing to secure the transfer of data among different CDCs using **IoT** devices.
2. Decentralization and virtualization chain of data is required for **blockchains** to work in SDN paradigm.
3. There is a need to deploy SDN-based cloud computing environment using **AI** learning models, which can reduce energy consumption and increase reliability while providing the network virtualization service in a cost-effective manner.

**4.8 Big Data Analytics and Data Science**

A complicated procedure of examining large datasets to uncover hidden patterns [33], market trends, correlations and preferences specific to customers that can help the companies to make well informed decisions [127]. In simple words, these technologies help in analysing data sets and then to draw conclusions from it using cloud computing platform. It is a form of analytics which involves elements as statistical algorithms, predictive models etc. Driven by high computing powered systems, it offers several advantages including effective marketing, revenue opportunities, operational efficiency etc [43] [44] [45]. It allows professionals to analyse the growing volumes of unstructured, semi structured and structured data. There are few research directions in this area as discussed below:

*4.8.1 Healthcare:* Large amount of data is generated in the healthcare industry [34] ranging from medical health records, X- ray reports, diet regime, record keeping etc. In order to give efficient cloud services, it's necessary to analyse this healthcare ecosystem-based data. Also, there is a need to build a fog, edge or cloud-based system for real time analysis on the enormous data set (collected by IoT devices). Including this, there is a need to keep this data tamper proof using blockchain models. Further, future research directions can be:

- *Patient Services:* Big data analytics-based systems can provide evidence-based medicines, giving faster relief to the patients as they can detect diseases at the earlier stages based on the clinical data available. This will help in minimizing drug doses to avoid side effect and reducing readmission rates thereby reducing cost for the patients. Customized patient treatment could be delivered by monitoring the effect of dosages of medicine continuously and looking on analysis of the data generated by the patients who already suffered from the same disease using cloud computing system.
- *Detecting Diseases:* Viral diseases can be predicted earlier before spreading, based on the real time analysis. This can be identified by analysing the history of the patients suffering from a disease in a particular geo-location [39]. This helps the healthcare professionals to advise the victims by taking necessary preventive measures.
- *Hospital Management:* Hospital's inventory could be planned and managed in advance to tackle problems like seasonal demand, uncertainty and economies of scale.

*4.8.2. Government:* Any government of the nation also generates petabytes of data [58]. Big data-based systems can assist government in providing value added services to its citizens. These systems could help the government in financial, healthcare, education budget planning by understanding the data patterns and the relationship between them with the help of machine learning algorithms [35] [63]. Further, future research directions can be:

- *Unemployment:* Analysing the market conditions and data of students before, government can predict the jobs. Cloud computing-based system enables government to create curriculum for trainings in order to absorb youth in the different domains and organizations.
- *Decision Making:* By analysing the sentiments and predicting the future trends, government can improve the quality and speed of decision making [38]. The government could take advantage of big data-based cloud computing systems in understanding current conditions and acceptability of the society before taking any action. It will help in creating more acceptability of the government in citizens.

*4.8.3. Retail:* With new sources of data like social media, geo location sensor data (IoT or edge devices), it has created more opportunities for retail companies to get competitive advantage and unprecedented value. Cloud based Big Data analytic systems can make best decisions flows, uncover hidden patterns and understand customer behaviour. To better understand the value of big data analytics in the retail industry [36], let's take a look at the following use cases, which are currently in production in various leading retail companies.

- *Conversion and Campaign:* Customers today interact more than they were before, and these interactions are happening on new platforms like social media. So, retail companies can get holistic view of customers and understand their preferences [40]. Data Science and engineering is capable of correlating customer purchase histories and profile information, with their behaviour on social media sites. And these relations can reveal unexpected insights, in turn helping the retailer is likely to have higher conversion rates and reductions in customer acquisition cost. Using data science platforms, retailers can:
  ➢ Analyse the impact of different promotional campaigns on customer behaviour.
  ➢ Use customer purchase history to identify the needs then generate personalize promotions catering to customer's needs.

- ➢ Monitor customer social media activity to make timely offers to customers to incent online purchases.
- *Customer Churn Prediction:* Data-driven customer insights are critical to tackle challenge of customer churn prediction [41] [42]. It is done by predicting future churn from data of the past. Retailers can look at characteristics of customers that have churned before in order to predict something about current customers.

*4.8.4. Operational Analytics:* Improvement in complicated product life cycles cause retailers to employ big data-based technologies to deploy product distribution strategies to reduce time and costs associated with them [43]. The key to utilizing data science and cloud computing platforms is to increase operational efficiency by unlocking insights buried in sensor and machine data through machine learning and pattern recognition techniques. These analyses help in predicting trends, patterns and outliers that can improve decisions [37] and save millions of dollars in computing world.

*Open Challenges and Trends:* The interested readers can further explore using extensive surveys on Big Data Analytics [57] [66] and Data Science [169]. The open challenges and future research directions for *Big Data Analytics and Data Science* are summarized as follows:

1. There is a need of bio-inspired based big data analytics mechanisms to process the data of edge devices of **IoT** applications at runtime.
2. Efficient **blockchain** data structures need to be developed for efficient storage and retrieval of large amounts of data.
3. Cloud based Big Data analytic systems can utilize the **AI** based techniques to make best decisions flows, uncover hidden patterns and understand customer behaviour.

**4.9 Data Processing**

Before diving deep into data processing, Let's try to understand "What is the need of data processing in today's world? ". The present world is overwhelmed with information from various sources such as IoT devices, social media, smartphones (IoT or edge devices), medical health records, click stream data, ecommerce etc. According to DOMO's research, "By 2020, it's estimated that 1.7MB of data will be created every second for every person on earth" [103]. Imagine, with current 7.7 billion population of the world, 13 petabytes of data will be created per second leading to 1 Million petabytes a day. So, with this rapid generation of data, organizations are endeavoring to find the best tools to deal with this raw data and make sense out of it.

The processing of data is central to any data-related problem. It is one of the most interesting, time consuming phase of an analytics project. Nearly 70-80% time of a data analyst/scientists is spent in cleaning and processing the data to make it usable for any kind of statistical modelling. In simple terms, data processing is basically the collection and manipulation of data to get useful information out of it, which then can be used for analytics, business intelligence, machine learning, deep learning and reporting purposes etc. The processing of data can incorporate anything from collection, reporting, aggregation, summarization, validation, structuring the unstructured data or vice versa etc. Data can be of any kind like time-series, images, videos, textual etc. Depending on the size of data, processing can be done on a single core machine to multi-core or on cloud and GPU servers [104]. The processing of big data can be broadly classified into three categories:

- *Batch processing* - an efficient technique to process large amounts of data collected over a period of time from various IoT or edge devices.
- *Real time processing* - deals with a continuous stream of data inputs and involves processing of data in near real time i.e. with minimal latency and maximum security (blockchain).
- *Hybrid processing* - takes the volume aspect of batch processing and velocity aspect of real time processing and it is useful in applications that require analysis of huge volumes of static along with streaming data.

Apache Hadoop (a framework that allows distribution of large data processing across various connected computers using MapReduce programming model) [105] and Apache Spark (a unified analytics engine with in-memory data processing capabilities and having built-in modules for SQL, machine learning, streaming and graph processing) are the two main open source tools that are widely used across industry for the processing of big data [63] [128]. A lot of other data processing tools for specific data types, tasks are also available in the market but these two precisely dominate the industry.

With organizations investing heavily on Advanced Data Analytics and the growth of data in terms of volume, variety and velocity increasing, it is becoming expensive and demanding for organizations to scale on-premises infrastructure [57]. As a result of this, cloud is becoming a natural choice for these organizations for storage and processing of data. More and more companies are moving towards cloud services being offered by the big tech giants like AWS (Amazon Web Services) by Amazon, Microsoft Azure, IBM Cloud, GCP (Google Cloud Platform) and Google for data processing [61]. These cloud providers have wide variety of tools to manage, compute and do analysis on data depending upon the volume, variety and velocity of data one has. So, these cost-effective cloud services are not only providing organizations with advanced tools for faster data processing but are also handling the agility and scalability aspects of big data and thus leading to revenue growth [66].

The management of data and extraction of knowledge are two important parts of grand organizations and business companies. The speed of generation of data at both user and system end leads to various research issues in both research community and industry [59]. The infrastructure uses to manage data is growing swiftly as collected from IoT or edge devices, which leads to formation of large cloud data centers (CDC) [58]. The various flexible data management models (NoSQL/relational) are using in CDCs to handle the current data requirements. Further, modern large CDCs are more susceptible to failures and needs effective fault tolerance technique for effective management of data within CDC [63]. Moreover, IoT and scientific applications are increasing which further needs effective data management mechanism within large scale distributed system. Big Data and Deep Mining models motivated from AI and machine learning techniques can be used for effective analysis of large-scale data.

*Open Challenges and Trends:* The interested readers can further explore using extensive survey on Data Processing [170]. The open challenges and future research directions for *Data Processing* are summarized as follows:

1. **IoT** and scientific applications are increasing which further needs effective data management mechanism within large scale distributed system.
2. To ensure data is protected, **Blockchain** technology has been adopted in the IoT domain and other real time systems.
3. **AI** provides a lucrative avenue to optimize large systems with huge amounts of data with engineering simplicity and efficiency by allowing automated decision making instead of human encoded heuristics which provide more efficient decisions very quickly.

**4.10 Application Design**

It is estimated that 50 billion devices will be online and 40 % of the world's data will come from them with total expenditure of $1.7 trillion by 2020 [58]. This exponential growth of Internet based smart devices and IoT applications such as healthcare services, real time traffic control systems, precision agriculture, smart cities etc require faster processing, data storage and privacy along with secure and reliable communication [46] [59] [61]. Also, as the data generated by these devices are used to solve real time problems, integrity, consistency and availability of data must be guaranteed. Designing these complex applications for IoT devices is a challenge in itself. So, we need to come out with application designs/architectures that are not only scalable to handle humongous amount of data from these devices but also reliable and fast enough to give efficient performance [135]. So, following are the major concerns that need to be taken care of while designing these applications with cloud infrastructure.

- *Latency:* Time taken by a data packet for a round trip from IoT devices to cloud and back. It's a big concern for time sensitive data as a millisecond can make a huge difference leading to unwanted results [46]. For e.g., disaster sensing device fires alarm after the occurrence of a disaster won't solve the problem. Extremely time sensitive data should be analysed very near to the data source to provide response in near real time.
- *Bandwidth:* If all the data generated by these devices are sent to cloud for storage and analysis, then the traffic generated by these devices will be simply gigantic and will consume all the bandwidth, which is not desirable. Also, as the physical distance between the device and cloud increases, transmission latency increases with it, increasing response time and stressing out the user. So, some work needs to be offloaded from the cloud, which can be done by allowing some processing to be done on an edge server that is positioned between cloud and device and physically closer to the device.

The fog computing allows IoT data storage and some processing locally at IoT devices and thus, avoids an excessive exploitation of Cloud resources [46]. Also, the fog provides reliability to time-sensitive and data-intensive applications that are large-scale and geospatially distributed [6]. Subsequently, fog computing might be viewed as the best decision to empower the IoT to give reliable and secure services/resources to numerous IoT users.

Big Data Analytics, IoT devices, fog and edge computing are becoming the driving forces for smart city initiatives throughout the world. Fog computing has great applicability in transportation such as vehicle to vehicle communications, managing smart sensor-based traffic control systems and also controlling autonomous vehicles, self-parking etc [46] [128]. It is also a more sustainable approach due to its low energy usage, small footprints and governments in various countries can use these applications to make the life of the citizens more secure and environment friendly. It can also be used in emergency services like fire, natural disasters by early notification of emergency situations to support smart decision making.

Farming applications help to oversee agriculture data like precipitation, wind speed and temperature to improve the use of climate and land in a productive way, which can assist farmers to have a productive yield [59]. An IoT agriculture platform for cloud and fog computing is proposed in [101], which can be utilized for pest management image analysis and monitoring, agricultural monitoring automation etc that can help farmers in better utilisation of resources. In [102], authors have proposed a fog computing application for precision agriculture that can assist in agricultural land management using AI based intelligent systems.

Also, it is progressively penetrating into the healthcare domain [61]. A lot of wearable gadgets like fit bit, blood pressure and heart rate monitoring cuffs, are being used to monitor different parts of human body and also collect information for diagnosis and interpretation. These devices have made remote healthcare monitoring feasible and hence doctors can monitor patients' wellbeing remotely and for the most part has given patients more authority over their lives and treatment. Also, companies like Apple with CareKit, HealthKit, ResearchKit and Google with Google fit, Microsoft building their health data management on top of Azure are clear examples that tech giants are investing heavily into digital healthcare [6] [7] [11].

*Open Challenges and Trends:* The interested readers can further explore using extensive survey on Application Design [135]. The open challenges and future research directions for *Application Design* are summarized as follows:

1. How to design a new application for smart cities to manage **IoT** based data effectively?
2. Storage capacity and scalability of applications are highly debated due to high cost and maintenance overheads while providing **blockchain** based security.
3. **Artificial Intelligence** algorithms can be used for processing of application data collected from various IoT based applications such as healthcare, agriculture, smart home etc.

**4.11 Serverless Computing**

Cloud application is basically comprising of three components: application logic, business logic and database server [6]. To improve the design of existing cloud applications, serverless computing paradigm is emerged [135]. In serverless application, database server and application logic located in the cloud while business logic is forwarded to the end user, which can be accessed by using web or mobile application for execution on provisioned resources without renting the resources (VMs). With the help of serverless computing, the different research challenges such fault tolerance, load balancing and under or over provisioning of resources are solved [58]. Further, serverless computing also decreases the coding part of developers and reduces the burden on cloud administrator for management of resources. Serverless provides two different kinds of service: 1) Function as a Service (FaaS) and Backend as a Service (BaaS) and these services are supported by Amazon AWS, Google Cloud and Microsoft Azure [128]. Cloud user only runs their application without knowing the internal details about servers, which are managed by cloud provider.

Serverless computing comes with many challenges and issues. Most works do not consider various aspects important for scheduling tasks on such execution models. One such aspect is server start-up time for infrequently used applications where servers are spun-down when the application is not in use. This severely affects the performance of application and QoS. Recent works like [19] do focus on this aspect but from a limited perspective. Another aspect is bandwidth consumption in Bag-of-Tasks models where multiple tasks commonly share files. These files need not be uploaded to cloud nodes separately for each task and task placement can be more intelligent to maximize file sharing capability. Yet another aspect important in such models is the security and privacy of applications and critical data [20]. As mentioned in [21], most modern serverless computing models are being implemented by integrating the edge of the network. These edge devices are resource-constrained and cannot support the heavy security applications and firewalls developed for common personal computers. Special applications and algorithms need to be developed to allow more secure communication as well as ensure the privacy of data for modern computing platforms which include edge devices as part of the datacentres.

*Open Challenges and Trends:* The interested readers can further explore using extensive survey on Serverless Computing [171]. The open challenges and future research directions for *Serverless Computing* are summarized as follows:

1. New **IoT** based applications are required to be developed to allow more secure communication as well as ensure the privacy of data for modern computing platforms which include edge devices as part of the datacentres.
2. Edge devices of IoT application are resource-constrained and cannot support the heavy security applications and firewalls developed for common personal computers, so there is a need to implement **Blockchain** technology to improve security.
3. **AI** systems improve the design of application for serverless computing.

**4.12 Deep Learning**

Deep learning is a class of machine learning programming technology that is based around learning from large data sets [154] [155]. The core of deep learning is to get high level interactive features from the raw data. Lately deep learning has been powering Reinforcement Learning to help realise the field of Deep Reinforcement Learning which is offering hope in crafting better models in the future [129]. The intersection of deep learning and cloud computing is creating interesting applications wherein both the fields are complementing each other. We would first discuss how cloud computing is supporting data scientists and later move on to how deep learning has been leveraging solutions to various traditional problems in cloud computing [49] [50] [51].

Teerapittayanon et. al [48] have demonstrated the use of edge and IoT end devices in realising a distributed training of a deep neural network. Traditionally distributed learning has been avoided due to communication cost, however they propose joint training bringing this cost down by 20x. Cloud computing is also helping reduce the energy consumption of deep neural networks by reducing the feature size achieved through splitting the network architecture between mobile and cloud [54]. There are various other applications of cloud computing assisted deep learning leading to value addition in various fields including robotics, autonomous driving, healthcare, personal assistance and defence. The advances in deep learning have to be realised by performing continuous training on the real time data. Cloud Chaser [60] is a technique in which the computational load is borne by the cloud and the power of deep learning can be harnessed on low computing capacity devices. Big Data Analytics can be effectively performed by employing a hybrid "Machine Learning + Cloud Computing" approach [57]. Hence a cloud-based deep learning approach is becoming very popular. Recent advances in deep learning and its wide popularity has led to a demand for learning the models on devices presently accessible to the users. A privacy preserving approach wherein on-device deep learning can be realised using cloud has been developed by [62]. Similar work has been carried out by various other groups trying to make deep learning on mobile devices a reality [50].

Nguyen et. al [49] have developed a deep learning model to thwart cyber-attacks in the context of mobile cloud realm. A Hybrid approach of running deep neural network partially over an IoT device and the cloud can lead to preserving the privacy over mobile devices [51]. They show a high accuracy (95.84%) with robustness using variety of datasets. Security has traditionally been the reason hindering adoption of cloud computing by users which can be addressed using deep learning as shown by many authors recently A persisting problem with cloud computing is managing and monitoring of large cloud clusters. Stefanini et. al [70] have shown that deep learning (DeepConv and DeepFFT) can be used to classify clusters with similar behaviour which may improve the scalability of managing a data centre. QoS violations can be detected ahead in time by "Seer" which is a cloud-based debugging system using deep learning and spatial and temporal data of cloud systems [64]. The prediction of workload for VMs on cloud has been carried out by [65] [68] [69]. Li et. al [71] have developed a system using cloud computing and deep learning with the aim of minimizing power consumption by cloud clusters.

*Open Challenges and Trends:* The interested readers can further explore using extensive survey on Deep Learning [172]. The open challenges and future research directions for *Deep Learning* are summarized as follows:

1. An ensemble deep learning based smart healthcare system is required for automatic diagnosis of heart diseases in integrated **IoT** and Fog computing environments.

2. Deep learning-based techniques are required to improve the **blockchain** structures.
3. How to enable deep learning on IoT devices to improve real world performance in **Artificial Intelligence** based intelligent systems?

**4.13 Cloud Containers**

Container technologies is quite popular in this cloud computing community with the origin of Dockers to execute the user workloads in an efficient manner. Container technology offers a lightweight cloud environment to deploy applications because containers are self-contained and stand-alone, which can reduce data dependency among various units during the execution of user workloads [130]. Container allows the resource sharing among various applications while running in an isolation manner. The various kernel features of Linux OSs such as libcontainer and control groups (cgroups) are considered in container technology. Namespaces and cgroups are used by Docker to execute self-dependent containers within a physical node and offer run resources (network, storage, memory and processor) in an isolated manner. Moreover, namespace separates the view of an application in running environment and simplifies the application deployment and increases the implementation efficiency. Further, container technology becomes a benchmark to develop, publish and execute applications in an isolated manner and denoted as a Container as a Service (CaaS) [130]. CaaS has three main advantages [63] [47]: 1) container starts very rapidly and takes less than a second to launch, 2) it uses very less amount of resources as compared to VMs and 3) container technology allows to run more instances concurrently.

Based on the current research in container technology, there are few research challenges are open which can be addressed in the future. Firstly, there is a weakness in security of containers due to the sharing of kernel as compared to VMs, which can be improved in the future by developing new security mechanisms by using Unikernel. Secondly, the performance improvement of containers is long task and there is a need to consider slack time to optimize the performance. Finally, there is a need to manage the user's QoS based container clusters using emerging cloud computing technologies such as Swarm and Kubernetes.

*Open Challenges and Trends:* The interested readers can further explore using extensive survey on Cloud Containers [173]. The open challenges and future research directions for *Cloud Containers* are summarized as follows:

1. Container technology can be utilized a cloud environment to deploy **IoT** applications to reduce data dependency among various units during the execution of user workloads.
2. there is a weakness in security of containers due to the sharing of kernel as compared to VMs, which can be improved in the future by adapting **Blockchain** technology.
3. There is a need to manage the user's QoS based container clusters using emerging **ensemble machine learning** techniques such as Swarm and Kubernetes.

**4.14 Quantum Computing**

Quantum computing is branch of computing in which we harness and exploit the laws of quantum mechanics to process the data. Classical computers cannot solve problems of certain size and complexity. Quantum computers leverage the phenomena of superposition and entanglement to solve the hard problems [136]. Superposition is phenomena by which quantum computer can be in multiple states at the same time. While entanglement is the phenomena of having strong correlation between two or more quantum particles that they are inextricably linked, even if they are separated by large distance. Due to these two principles, a quantum computer can do large number calculations simultaneously [38]. Think of it this way: whereas a classical computer works with ones and zeros, a quantum computer will have the advantage of using ones, zeros and "superpositions" of ones and zeros [137]. Certain difficult tasks that have long been thought impossible (or "intractable") for classical computers will be achieved quickly and efficiently by a quantum computer.

The future directions [136] [137] [138] [156] for this research area are:

*1. Optimization:* Consider an example of management company, which wants to invest in large cap value, medium cap value and blue-chip companies to generate large returns while rebalancing the asset classes to protect investments. By combining dynamic asset allocation through the use of quantum computers, they can solve this kind of hard problem optimally. These kinds of problems exist in various domains like airline traffic scheduling, financial analysis, system design etc. they are the one of the most complex problems [40] of the world with potential to transform the lives of the people. Quantum computers will be able to calculate the one-way functions, including blockchains, that are used to secure the Internet and financial transactions.

*2. Machine Learning and AI:* Quantum computing can transform field of machine learning field too. Object detection in which it is very easy for humans to pick different objects from the photograph. But for traditional computers it is a difficult task. As the programmers don't know how to write code that can infer many objects by its own. Machine learning is one of the approaches to solve this problem in which algorithms recognize the objects by getting trained on large datasets [41]. As the amount of data and its combinations involved in this process are very large. So, it becomes a computationally expensive problem for the traditional systems to handle. Quantum computers make these types of problems easy as they can do enormous calculations simultaneously.

*3. Material Simulation:* Quantum computers make simulating the materials feasible, the material simulation field can lead to development in various IoT based application in robotic, chemical and optical industry [39].

*Open Challenges and Trends:* The interested readers can further explore using extensive survey on Quantum Computing [174]. The open challenges and future research directions for *Quantum Computing* are summarized as follows:

1. Quantum computers make simulating the materials feasible, the material simulation field can lead to development in various **IoT** based application in robotic, chemical and optical industry.
2. Quantum computers will be able to calculate the one-way functions, including **blockchains**, that are used to secure the Internet and financial transactions.
3. Quantum computing can transform field of machine learning field too. Object detection in which it is very easy for humans to pick different objects from the photograph. As the programmers don't know how to write code that can infer many objects by its own. **Machine learning** is one of the approaches to solve this problem in which algorithms recognize the objects by getting trained on large datasets.

**4.15 Bitcoin Currency**

Blockchain was originally developed for digital currency Bitcoin and proposed as solution for settlement of transactions [106]. The blockchain is an incorruptible chain of data blocks validated with PoW of economic transactions that can also be programmed to record not just for financial transactions but virtually everything of value [107]. Blockchains, like Bitcoins and Ethereum offer a new paradigm to run distributed applications. The developers define smart contracts for Bitcoin currency transactions, and the contracts are executed on the blockchain virtual machines. Therefore, the blockchain uses a distributed runtime environment with distributed consensus.

The Bitcoin supporting network also allows block of data to be distributed across ledgers via peer to peer network without central management. Anyone can join the network, and the data in blockchain is validated by the participants to make the data secure and open. This feature can be taken advantage by cloud computing, especially for the security of cloud storage.

Cloud infrastructures provide computing power to run large applications and process huge amount of data. However, to manage the huge data storage, the centralized data centers connected with the Fog or IoT devices at the network edge cannot offer an efficient way to provide high availability, real-time and low latency services [108]. To address these issues, distributed cloud architecture rather than the traditional network architecture is needed.

Blockchain offers some necessary features to build a distributed cloud [109] [110] [111], such as: 1) It can facilitate resource usage via distributed applications to enable fine-grained control on resources. 2) QoS can be improved as blockchain can provide traceable resource usage, thus the user and service provider can verify whether the QoS is ensured. 3) A market place that everyone can advertise their computing resources and find the needed resources using AI based techniques or prediction models.

Comparing to cloud computing, blockchains only have limited computing resources to execute distributed applications, e.g. limited storage, inefficient virtual machines and protocol with high latency. Therefore, for the latency-aware applications and resource-intensive applications, these issues should be overcome. Combining blockchain and cloud together to establish the blockchain-based distributed cloud can bring new benefits and overcome existing limitations. The blockchain-based distributed cloud allows on-demand resource, secure and low-cost access to infrastructure, the data is also brought to be closer to the owner and consumer [15]. Meanwhile, blockchain-based distributed cloud is possible to overcome the expensive and high energy consumption features of clouds. Another future direction that blockchain can benefit is improving the security of cloud storage. User data can be split into small blocks and added one more security layer, then the small blocks can be stored in distributed locations. The hacker can only get a chuck of the data rather than the whole file. The hackers who want to alter the data can also be removed from the network, and the altered data can be recovered from redundant copy.

*Open Challenges and Trends:* The interested readers can further explore using extensive survey on Bitcoin Currency [175]. The open challenges and future research directions for *Bitcoin Currency* are summarized as follows:

1. How **IoT** botnets affect the "internet of money" cryptocurrency?
2. The **blockchain** is an incorruptible digital ledger of economic transactions that can also be programmed to record not just for financial transactions but virtually everything of value. **Blockchains**, like Bitcoins and Ethereum offer a new paradigm to run distributed applications.
3. There is a need to investigate how **Machine Learning (ML)** techniques perform in the prediction of cryptocurrency prices.

**4.16 Software Engineering**

Software Engineering (SE) and cloud computing are very close paradigms. For example- Service-oriented SE combines the best features of the services and cloud computing and thus gives several benefits to software development process and applications. The only difference between Services oriented SE and cloud computing is that –the service-oriented SE focuses on architectural design (service discovery and composition) and cloud computing focuses on the effective delivery of services to users through flexible and scalable resource virtualization and load balancing [163]. Software Engineering not only evolves hardware technologies, but also involves customers and software developers [112]. Cloud computing and virtualization allow users to create VMs and cloud services for projects and software's with automatic management [113]. With cloud services, the software development teams can combine development, test and delivery processes seamlessly. Cloud computing can enhance the software engineering process in the following ways:

1. The development process can be speed up. Cloud computing and virtualization offer sufficient computing resources, so that developers can use multiple virtual machines rather than stick to a single physical machine.
2. Cloud computing enables the development activity into a more parallel way, as the time to install the necessary applications can be reduced by fetching cloud services, which can lead to a more efficient development process.
3. Cloud instances and virtualization can greatly enhance the integration and delivery process. With adequate virtualization resources from their own cloud or public cloud, development can make the build and test process faster, which are quite time-consuming.
4. Code versions management becomes easier. Code branching is necessary in the code refactoring or function increment in software development. Cloud computing relieves the efforts to buy or rent physical machines for storing the codes.
5. Cloud environment provide interfaces to facilitate users' access to applications and can improve service QoS via dynamic resource provisioning.

In conclusion, cloud computing removes the heavy dependencies of development servers on fixed physical machines and makes the development process in software engineering more efficiently [114].

Cloud computing makes the software development process to be more efficiently. However, some challenges exist in combining software engineering and cloud computing, and these challenges should be addressed. Data migration is one of the challenges. Since cloud providers offer different APIs for offering cloud services, if the software's and data need to be migrated to another cloud, software's may need to be reimplemented and some software's can be the legacy system [115]. To address this issue, when developing and deploying software's in clouds, unnecessary dependencies on specific APIs should be avoided. Another challenge is the reliability and availability. If all the data are migrated to clouds from local, when the cloud is attacked by hackers or influenced by unpredicted disaster, the data is hard to recover. This requires the developers to prepare local backup.

Cloud computing provides new possibilities for software engineering researchers to explore multilateral software development [2]. Several researchers attempted to use cloud computing for reducing the cost of operation, delivery and software development [164]. In [164] authors explored how to replace Learning Management Systems (LMS) services with a cloud platform for sharing the knowledge and collaboration among university students. Software systems are being replaced by cloud-based systems to save the cost and maximum utilization of resources.

In recent years when data is increasing on exponential rate then it is not easy to use the old traditional way of handling the data. The emerging technologies based on IoT, blockchain, machine learning, and artificial intelligence are opening a new area of research in software engineering and the major issue in these technologies are handling the huge amount and variety of data. These researches also giving the opportunity to new researches and the new ways of handling the data in the cloud and this results for starting of upgraded technologies like- Fog Computing- first used by Cisco in order to extend the current cloud computing infrastructure [165]. Software companies working for building enterprise software are building an abstraction layer and offering it as a service, called Blockchain-as-a-Service [165]. All these emerging areas are new but heavily depend on software engineering.

*Open Challenges and Trends:* The interested readers can further explore using extensive survey on Software Engineering [176]. The open challenges and future research directions for *Software Engineering* are summarized as follows:

1. To investigates the feasibility of **IoT** based Software engineering solutions on how organizations can deliver high business value through technology and operations strategy engagements at the same time can generate Return On Investment (ROI) by effectively utilizing the possibilities of IoT in business.
2. There is a need for software engineers to devise specialized tools and techniques for **blockchain**-oriented software development.
3. How to program the system "automatically" using **Artificial Intelligence** instead of writing the software code manually?

**4.17 5G and Beyond**

Next generation networks promise not only extremely high data rates and low latency, but also ubiquitous coverage and massive IoT. Today, dense network deployment is one of the effective strategies to meet the capacity and connectivity demands of the fifth-generation (5G) cellular system [117]. 5G is the fifth generation of cellular mobile communications, preceded by 4G (LTE/WiMax), 3G (UMTS), and 2G (GSM) systems. 5G are expected to provide high-speed data rates, reduced latency, reduced energy consumption, effective cost, enhanced system capacity, and massive simultaneous device connectivity [118]. One of the main objectives of 5G networks is to support applications that involve a high density of devices. In this regard, the concepts of massive Machine-Type Communications (mMTC), enhanced Mobile BroadBand (eMBB) and Ultra-Reliable Low-Latency Communications (URLLC) are being developed to support such applications [119]. 5G will bring major advances in the network and communications systems by providing ultra-high-speed data transmission that can be 100 times faster than the existing 4G [120].

Edge computing as an evolution of cloud computing shifts the application hosting paradigm from the centralized data centres to the network edge, closer to consumers and the data generated by applications [121]. Edge computing is considered one of the main enablers for satisfying the demanding Key Performance Indicators (KPIs) of 5G such as enhanced mobile broadband, low latency and massive connectivity. The boosting computation-intensive applications in the Internet of Things (IoT) era along with the growing number of mission-critical tasks in emerging networks is a main bottleneck in the design of the real-time communication systems [122], [123]. In order to tackle the massive computing demands and the scarcity of the resources (i.e., small size and low power) available at the mobile device, Mobile Edge Computing (MEC) is considered as a promising solution to enhance mobile user's computation capability and realize low-latency communications [124]. Cloud-like MEC server along with the Access Point (AP) at the edge of networks [125]. The advantage of MEC enables the resource-limited mobile users to offload tasks for remote execution at the more powerful MEC server in their proximity, which brings the benefit of improved computation capacity and reduced latency [126].

The advent of 5G and cloud computing will enhance the capacity, functionality, and flexibility of a network operators to offer new range of services [117].

*Open Challenges and Trends:* The interested readers can further explore using extensive survey on 5G and Beyond [177]. The open challenges and future research directions for *5G and Beyond* are summarized as follows:

1. The boosting computation-intensive applications in the **IoT** era along with the growing number of mission-critical tasks in emerging 5G networks is a main bottleneck in the design of the real-time communication systems.
2. 5G networks needs to use **Blockchain** technology to deliver secure communication.
3. Data analytics on massive amount of data collected from the massive number of sensors in the Industrial IoT use cases can be managed in cost effective manner by using 5G based **Artificial systems**.

**4.18 Autoscaling**

Elasticity feature of cloud computing has brought about the opportunity for utilizing self-adaptive solutions to minimize the cost of resources while preserving QoS. Self-additivity is realized through resource auto-scaling, aka planning, reconfiguration and provisioning. Auto-scaling, i.e. dynamically adjustment of computing resources like Virtual Machines (VMs), is a widely investigated mechanism [178], by which researchers mainly seek for a) horizontal adjustments, i.e., adding/removing VMs; b) vertical adjustments, i.e., adding/removing resources of VMs [158]; c) decision-making methods, including analytical modelling, control theory, and machine learning [1]; d) leveraging varied pricing model, i.e., On-Demand, Reserved and Spot [159]; and, e) substituting light-weight container-based machines with hypervisors [160]. Motivated by QoS requirements, particularly latency, auto-scaling mechanisms are always facing two main questions: how to reach a scaling decision timely? And, how to execute the decision timely?

To reach timely decision, first and foremost is to use forecasting using AI. However, conventional machine learning might be inefficient when it comes to IoT applications requiring timely error correction, as they lack automotive correction without human-intervention. Besides, while the need for timely execution and provisioning of resources, in seconds, in cloud was about to be accomplished by utilizing container-based solution and by providing burstable performance resource [161], latency-critical IoT applications and microservices requiring responds in the scale of millisecond appeared—deteriorating the situation. A smart car, for instance, continuously senses the wheel speed, the pedestrians' movement, and vehicle surrounding; if tailgating is about to happen, for example, it must decide in milliseconds and activate the brake-by-wire system, otherwise a disaster might happen. This challenge requires auto-scaling mechanisms for IoT applications to be aware of mobility, geo-distribution, and location, for which cloud is unable to provide solutions lonely because of unstable and long-delay links between cloud and users [156]. In fact, cloud naturally conflicts with Industry 4.0 (i.e. digitization of manufacturing) principles, e.g. real-time controlling and decentralized decision-making, hence auto-scaling needs extending.

Filling such gaps requires two main actions: a) collaboration between insular clouds and b) bringing the computing close to users/IoT applications—Fog/Edge computing. However, such collaborative environments, firstly, raise serious monetization concerns for cloud providers [162]. Secondly, although Fog/Edge can extend auto-scaling mechanisms execution so that they can be performed by devices in the close proximity to users or in fog nodes at the edge of network, as well as cloud, such decentralized environments demands the actual realization of 5G networks which is still in the infant stages of development. Further effort could be intelligent prediction and decision-making which is becoming the first-class citizens of every optimization problem; however, tradition machine learning models, for instance, return inaccurate predictions in some cases which require to be fixed by the programmers, conflicting with latency requirements of applications.

*Open Challenges and Trends:* The interested readers can further explore using extensive survey on Autoscaling [178]. The open challenges and future research directions for *Autoscaling* are summarized as follows:

1. Extending auto-scaling across the compute continuum from **IoT** devices to cloud in order for timely execution/processing of scaling decisions is a challenging problem that needs to be addressed.
2. Blockchain-enabled auto-scaling utilizing **Blockchain** capabilities such as Smart SCs in order for addressing first security concerns and second unsolved monetization problem in federated cloud for services provided in collaborative computing is a new challenge.
3. Self-correction prediction and multi-objective auto-scaling using **AI** for the trade-off between performance and cost can be accomplished using Deep Learning. Recently, deep learning, i.e. automation of predictive analytics—a subset of AI—has gain more attention for solving the problems which have not been yet solved.

## 5. Insights of Triumvirate to the Cloud Computing Evolution: A vision

The purpose of the study would be to see how the three new technologies (blockchain, IoT and Artificial Intelligence) will influence the evolution of cloud computing. We reviewed 140 research papers in this systematic review and presented them in a categorized manner and it comprises of most recent research work related to cloud computing paradigms and technologies. We discussed the research issues addressed and open challenges that needs more attention in future to conduct next level research. Figure 3 shows the insights of the triumvirate (blockchain, IoT and Artificial Intelligence) to the evolution of cloud computing.

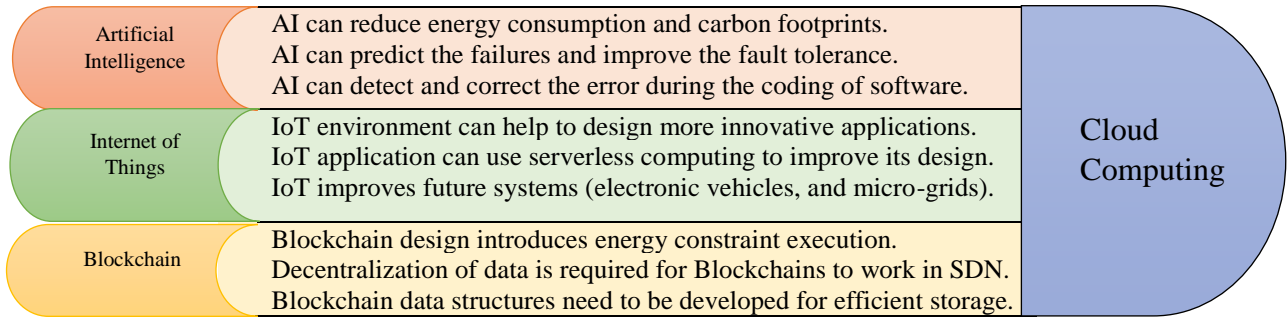

Figure 3: Insights of Triumvirate to the Cloud Computing Evolution

## 6. A Conceptual model for Cloud Futurology: Holistic View

To resolve the above-mentioned challenges, there is a requirement of conceptual model to explore the influence of three emerging paradigms (Blockchain, IoT and Artificial Intelligence) on evolution of cloud computing. Figure 4 shows the conceptual model, which describes the transformation effects of paradigms and technologies on cloud computing evolution. The model integrates the three paradigms: IoT, Blockchain and Artificial Intelligence to provide a holistic view of an abstract design encompassing multiple domains in computer science. Various components of IoT including sensors and actuators communicate with the gateway, all connected through 5G technology. The user interacts with the model using app interface on the gateway device with compute costing through Bitcoin (or related cryptocurrency) technology. The gateway layer also contains basic pre-processing data engine and various IoT applications for task generation and result collection in seamless fashion. The tasks are sent to and managed by the Broker nodes which authenticate payments and other transactions with sophisticated energy and resource management modules, backed with fault tolerance and security systems. These nodes are also connected to the database and external controllers through SDN. The computation tasks are executed on fog or cloud machines (physical or virtual) having robust data analytics, deep learning and blockchain mining capabilities. Software engineering and autoscaling allow end-to-end integration with cloud allowing resource availability from server-based deployments to serverless frameworks. The model is enhanced by quantum computing technologies. Overall, the model integrates and enables computation using a plethora of technological advancements and provides an enhanced and holistic setup for next generation computing environments.

## 7. Summary and Conclusions

Cloud computing is an emerging paradigm, enabling on demand, metered access to compute resources (Process, Memory, Storage, etc) driving technological innovation and enabling geographically distributed applications. In this review paper, we have presented the systematic review of computing paradigms and technologies and the influence of triumvirate (blockchain, IoT and Artificial Intelligence) to the evolution of cloud computing. The history and background of computing paradigms and technologies has been presented and designed its evolution. Further, the research areas related to cloud computing have been identified, discussed and the research issues and challenges are highlighted. We have proposed a conceptual model to explore the influence of three emerging paradigms (Blockchain, IoT and Artificial Intelligence) on evolution of cloud computing. We hope that this systemic review will be beneficial for researchers who want to do research in any area concerning to cloud computing.

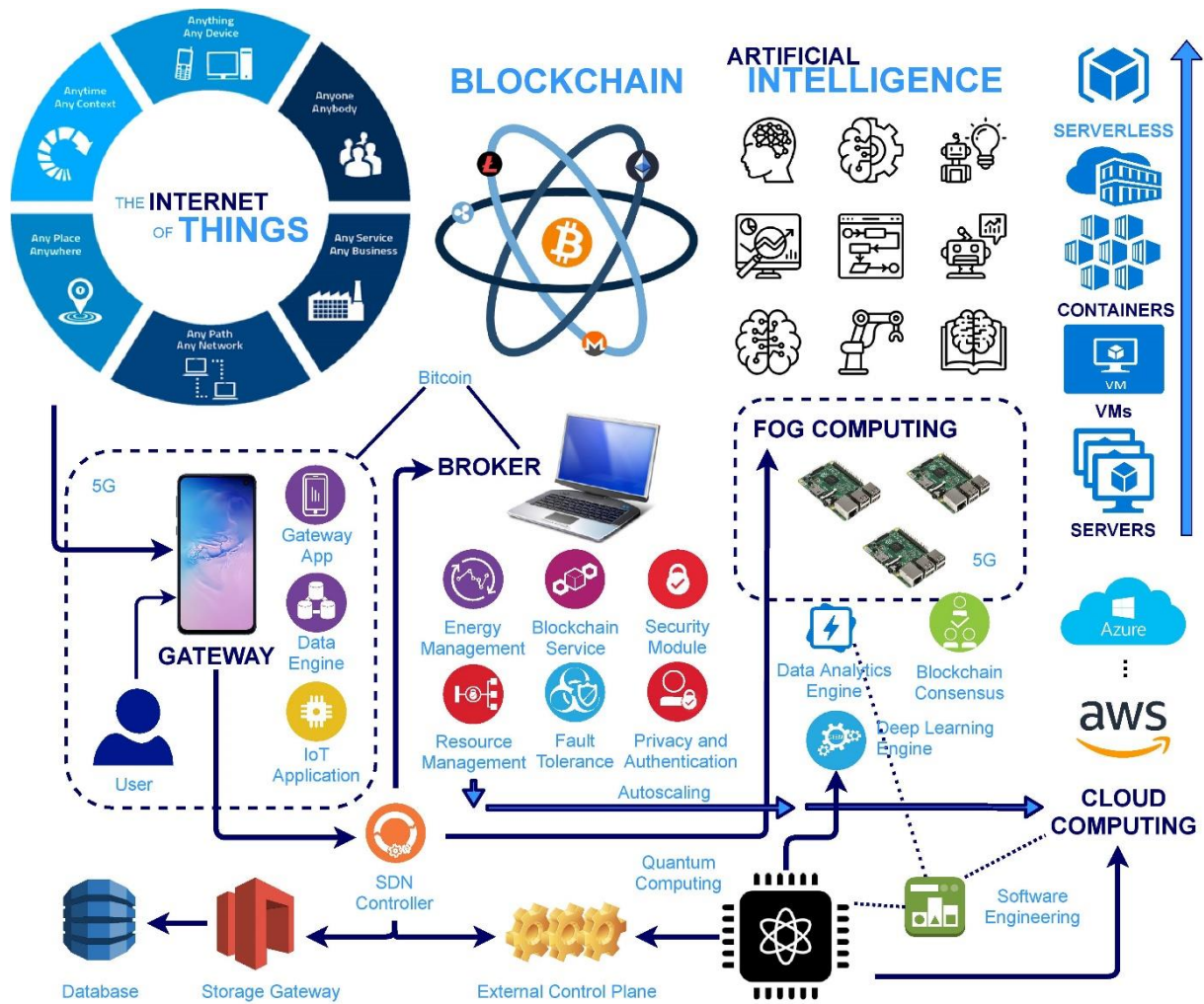

Figure 4: A Conceptual model for Cloud Futurology

**Declaration of Competing of Interest**

We do not have any conflicts of interest.

**Acknowledgments**

We would like to thank the editor, area editor and anonymous reviewers for their valuable comments and suggestions to help and improve our research paper.